\documentclass{article}
\usepackage[utf8]{inputenc}
\usepackage[margin=1in]{geometry}
\usepackage[affil-it]{authblk}
\usepackage{natbib}
\usepackage{amsmath}
\usepackage{amsfonts}
\usepackage{bm}
\usepackage[color=lightgray]{todonotes}
\usepackage[bf,font={small,sl}]{caption}

\title{\textbf{State-switching continuous-time correlated\\ random walks}}
\author{Théo Michelot$^1$\footnote{Corresponding author: tmichelot1@sheffield.ac.uk}, Paul G. Blackwell$^1$}
\affil{$^1$University of Sheffield, UK}
\date{}

\begin{document}
\maketitle

\begin{abstract}
\noindent
1. Continuous-time models have been developed to capture features of animal movement across temporal scales. In particular, one popular model is the continuous-time correlated random walk, in which the velocity of an animal is formulated as an Ornstein-Uhlenbeck process, to capture the autocorrelation in the speed and direction of its movement. In telemetry analyses, discrete-time state-switching models (such as hidden Markov models) have been increasingly popular to identify behavioural phases from animal tracking data.\\ 
2. We propose a multistate formulation of the continuous-time correlated random walk, with an underlying Markov process used as a proxy for the animal's behavioural state process. We present a Markov chain Monte Carlo algorithm to carry out Bayesian inference for this multistate continuous-time model.\\
3. Posterior samples of the hidden state sequence, of the state transition rates, and of the state-dependent movement parameters can be obtained. We investigate the performance of the method in a simulation study, and we illustrate its use in a case study of grey seal (Halichoerus grypus) tracking data.\\
4. The method we present makes use of the state-space model formulation of the continuous-time correlated random walk, and can accommodate irregular sampling frequency and measurement error. It will facilitate the use of continuous-time models to estimate movement characteristics and infer behavioural states from animal telemetry data.
\end{abstract}

\vspace{1em}
\noindent
{\bf Keywords:} animal movement, continuous time, Ornstein-Uhlenbeck process, state-space model, multistate model, random walk
 \vspace{1em}

\section{Introduction}
The collection of large high-resolution animal tracking data sets has motivated the development of a wide range of statistical methods \citep{patterson2017, hooten2017}. For the analysis of animal movement, an important conceptual modelling choice is the time formulation, i.e.\ the choice between discrete-time and continuous-time models \citep{mcclintock2014}. Although animals move in continuous time, their location may only be observed at discrete intervals (e.g.\ every minute or every hour). Dicrete-time approaches are based on the assumption that the underlying movement process can be appropriately modelled at the time scale of the observations. Most often, movement is described by the ``step lengths'' (distances between successive locations) and ``turning angles'' (angles between successive directions), derived from the location data \citep{morales2004, jonsen2005}. However, the distributions of these metrics of movement strongly depend on the sampling rate, such that the resulting inference is tied to a specific temporal scale \citep{codling2005, schlagel2016}. One of the consequences is that discrete-time methods require locations to be collected at regular time intervals through the period of the study. Many telemetry data sets are collected at irregular time intervals, for example with marine mammals which may only be observed when they surface. To use a discrete-time model in such cases, it is then necessary to interpolate the data points on a regular time grid, introducing approximation uncertainty.

On the other hand, continuous-time models consider that telemetry observations arise from a continuous movement process. As such, they can naturally accommodate different temporal scales, and irregular sampling rates \citep{patterson2017}. Various approaches have been used to model animal movement in continuous time, most of them based on diffusion processes. These include Ornstein-Uhlenbeck processes \citep{dunn1977, blackwell1997, blackwell2003}, Brownian bridges \citep{horne2007}, and potential functions \citep{preisler2013}. A very popular model is the continuous-time correlated random walk (CTCRW) introduced by \cite{johnson2008}, in which the velocity of the animal is formulated as an Ornstein-Uhlenbeck process. Through the velocity, this model incorporates autocorrelation into both the speed and the direction of the movement, similarly to discrete-time correlated random walks based on step lengths and turning angles. \cite{johnson2008} formulated the CTCRW as a state-space model, making fast inference possible through the Kalman filter, and made it available in the R package crawl \citep{johnson2017}. \cite{fleming2017} extended this implementation to a wider family of diffusion processes, including their ``OUF'' model of correlated movement around a centre of attraction. \cite{gurarie2017} recently reviewed the use of the CTCRW model for the analysis of animal tracking data.

Random walks have been used as ``building blocks'' for more complex, multistate, models. These state-switching models describe animal movements as the outcome of several distinct behaviours, e.g.\ ``foraging'', ``resting'', ``exploring'', based on the notion that the behavioural states of the animal differ noticeably in terms of some metrics of the movement, e.g.\ speed or sinuosity \citep{morales2004}. Although this idea has received a lot of attention for discrete-time models, with the growing popularity of hidden Markov models \citep{patterson2009, langrock2012}, it has been underutilised in continuous-time approaches. \cite{blackwell1997} introduced a multistate movement model, where the location of the animal is modelled with an Ornstein-Uhlenbeck process. That model does not directly capture the movement persistence in speed and direction, which makes its application limited for high-frequency tracking data. More recently, \cite{parton2017} described a multistate approach in which the speed and the bearing of the animal are modelled with diffusion processes, analogously to discrete-time models based on step lengths and turning angles. However, their method requires computationally-costly numerical approximation to reconstruct the movement path at a fine time scale, a disadvantage in dealing with large tracking data sets. \cite{mcclintock2014} presented a multistate analysis based on the CTCRW, but they constrained the state process to be constant over each time interval between two observations. Therefore, they do not carry out exact inference from the continuous-time model.

Alternatively, to circumvent the limitations of discrete-time models, \cite{mcclintock2017} suggested a two-stage approach based on multiple imputation methods. A one-state continuous-time model (such as the CTCRW) is fitted to the data, and a large number $m$ of possible realisations of the movement process are simulated from the model on a regular time grid. Then, a hidden Markov model is fitted to each realisation, to investigate the state-switching dynamics. The $m$ sets of estimates are pooled, such that the resulting model takes into account the uncertainty in the locations. Note that, since the realisations are generated without taking into account the possible behaviours, this is not fully equivalent to fitting a multistate CTCRW model.

Here, we extend the framework of \cite{johnson2008} to incorporate behavioural states directly into the CTCRW framework, with an underlying continuous-time Markov process. We present the model formulation and describe a Bayesian estimation method to infer hidden states and movement parameters in this framework. We investigate the performance of the method in a simulation study, and then use it to analyse a trajectory of grey seal (\emph{Halichoerus grypus}).

\section{Model formulation} \label{sec:model}
\subsection{Continuous-time correlated random walk} \label{sec:ctcrw}
The continuous-time correlated random walk (CTCRW) was introduced as a model of animal movement by \cite{johnson2008}. The underlying stochastic process was originally developed by \cite{uhlenbeck1930} to describe the movement of a particle under friction.

Denote $\bm{z}_t = (x_t, y_t)^\top$ the location of the animal at time $t$, and $\bm{v}_t = (v_t^x, v_t^y)^\top$ its velocity, linked by the equation
\begin{equation} \label{eqn:vel}
	d\bm{z}_t = \bm{v}_t dt.
\end{equation}

In the CTCRW model, the velocity of the animal is modelled by an Ornstein-Uhlenbeck process, defined as the solution of the stochastic differential equation
\begin{equation} \label{eqn:OU}
	d\bm{v}_t = \beta (\bm{\gamma} - \bm{z}_t) dt + \sigma d\bm{w}_t,
\end{equation}
where $\bm{w}_t$ denotes a Wiener process, $\bm{\gamma} \in \mathbb{R}^2$ is the mean velocity, $\beta>0$ measures the reversion of the velocity to its mean, and $\sigma>0$ measures the spread of the velocity around its mean. In practice, the mean velocity parameter $\bm{\gamma}$ is often taken to be zero, corresponding to the case where there is no systematic drift in the animal's movement \citep[although see][for an example of analysis with drift]{johnson2008}. Here, for simplicity, we consider that $\beta$ and $\sigma$ are scalar parameters, corresponding to the isotropic case, but they could be taken as matrices for a more general formulation \citep{blackwell2003, gurarie2017}. We will sometimes refer to the location process $\bm{z}_t$ as an integrated Ornstein-Uhlenbeck process, to indicate that its derivative (with respect to time) is an Ornstein-Uhlenbeck process.

This formulation is very convenient because it is possible to derive the transition densities of the velocity process $\bm{v}_t$ and of the position process $\bm{x}_t$ analytically. In the following, we assume $\bm\gamma=\bm{0}$. Using Itô calculus, it can be shown that, under Equations \ref{eqn:vel} and \ref{eqn:OU}, we have
\begin{equation*}
	\bm{v}_{t+\delta} = e^{-\beta\delta} \bm{v}_t + \bm\zeta(\delta),
\end{equation*}
and
\begin{equation*}
	\bm{z}_{t+\delta} = \bm{z}_t + \left( \dfrac{1-e^{-\beta\delta}}{\beta} \right) \bm{v}_t + \bm\xi(\delta),
\end{equation*}
for any time interval $\delta>0$ and, for either dimension $c \in \{ x, y \}$,
\begin{align}
	\zeta_c(\delta) & \sim N \left[ 0,\ \dfrac{\sigma^2}{2\beta} (1-e^{-2\beta\delta}) \right], 
    \label{eqn:zeta}\\
	\xi_c(\delta) & \sim N \left[ 0,\ \left(\dfrac{\sigma}{\beta} \right)^2 
    	\left(\delta + \dfrac{1-e^{-2\beta\delta}}{2\beta} - 
    	\dfrac{2(1-e^{-\beta\delta})}{\beta} \right) \right]. \label{eqn:xi}
\end{align}
The steps of the calculation are detailed in Appendix S1.

\subsection{Multistate model} \label{sec:multistate}
In this paper, we use the CTCRW as a building block for more complex and realistic movement models. Multistate models of animal movement have been developed to account for behavioural heterogeneity. In the most common formulation, a (discrete- or continuous-time) Markov process models switches between discrete ``behavioural'' states, on which depend the parameters of the movement process \citep{blackwell1997, morales2004}. Following this idea, we introduce a $N$-state continuous-time Markov process $(s_t)_{t \geq 0}$, characterised by its infinitesimal generator matrix,
\begin{equation}
	\label{eqn:generator}
	\bm{\Lambda} = 
    \begin{pmatrix}
		- \lambda_1 & \lambda_{12} & \cdots & \lambda_{1N}\\
        \lambda_{21} & - \lambda_2 & \cdots & \lambda_{2N}\\
        \vdots & \vdots & \ddots & \vdots\\
        \lambda_{N1} & \lambda_{N2} & \cdots & - \lambda_N\\
	\end{pmatrix},
\end{equation}
where $\forall i \in 1,\dots,N,\ \lambda_i = \sum_{j \neq i} \lambda_{ij}$. At any time $t \geq 0$, the discrete state $s_t$ takes one of $N$ values $\{ 1, \dots, N \}$, typically used as proxies for the behavioural states of the animal (e.g.\ ``foraging'', ``exploring''). The generator matrix is the continuous-time analogue of the transition probability matrix used in hidden Markov models, and its entries determine the state-switching dynamics. In particular, as a consequence of the Markov property, the dwell times in state $i$ follow an exponential distribution with rate $\lambda_i$.

We now consider that the parameters of the CTCRW model ($\beta$ and $\sigma$ in Equation \ref{eqn:OU}) are state-dependent, so that each behavioural state can be associated with a different type of movement. Using the notation introduced in Section \ref{sec:ctcrw}, the multistate CTCRW model is defined by
\begin{equation*}
  \begin{cases}
      d\bm{z}_t = \bm{v}_t dt,\\
      d\bm{v}_t = -\beta_{s_t} \bm{v}_t dt + \sigma_{s_t} d\bm{w}_t.
  \end{cases}
\end{equation*}

This can be viewed as a higher-dimension continuous-time Markov process composed of a continuous component (the location and velocity processes) and a discrete component (the discrete state process), as described e.g.\ by \cite{berman1994}. In the following, we develop a method of Bayesian inference for the multistate CTCRW model. 

\section{Bayesian inference} \label{sec:inference}

\subsection{Likelihood evaluation with Kalman filter}
\label{sec:kalman}
We present a method to evaluate the likelihood of the multistate CTCRW model, given a reconstruction of the underlying state process. \cite{johnson2008} implemented a Kalman filter to obtain the likelihood of movement trajectories in the single-state CTCRW model. With only minor changes, it can be extended to evaluate the likelihood of the multistate model, conditionally on the state sequence. The values of the state process are typically not known, and will thus need to be imputed. One method of reconstructing the state sequence is presented in Section \ref{sec:mcmc}. 

We consider a data set of observed locations, augmented with the times of the reconstructed state transitions. The locations associated with the transitions are generally not observed, and they are thus treated as missing data. We denote by $\{ \tilde{\bm{z}}_1, \dots, \tilde{\bm{z}}_n \}$ the augmented sequence of locations, possibly observed with measurement error, and $\{ t_1, \dots, t_n \}$ the associated times. We denote by $s_i$ the (imputed) behavioural state between $t_i$ and $t_{i+1}$.

Following \cite{johnson2008}, the model can be written as a state-space model, where the observation process is the animal's location (possibly observed with noise), and the continuous state process is composed of both the true location and the velocity. We denote by $\bm{\omega}_i = (x_i, v_i^x, y_i, v_i^y)^\top$ the continuous state vector at time $t_i$, $\Delta_i = t_{i+1} - t_i$ the time intervals, and $\bm{\xi}_i = \bm{\xi}(\Delta_i)$ and $\bm{\zeta}_i = \bm{\zeta}(\Delta_i)$ the stochastic terms of the transition densities for the location and the velocity, respectively (Equations \ref{eqn:zeta} and \ref{eqn:xi}, substituting $\beta_{s_i}$ for $\beta$ and $\sigma_{s_i}$ for $\sigma$).

Both $s_i$ and $\bm\omega_i$ are referred to as ``state'', in state-switching models and state-space models, respectively. Here, we combine both and, to avoid confusion, we will refer to $\bm\omega_i$ as the ``continuous'' state of the model (as opposed to the ``discrete'' behavioural state $s_i$).

The observation equation of the CTCRW is
\begin{equation*}
	\tilde{\bm{z}}_i = 
    \bm{Z}
	\bm{\omega_i} + \bm\varepsilon_i,\qquad
    \bm\varepsilon_i \sim N(\bm{0}, \bm{H}_i),
\end{equation*}
where $\bm{H}_i$ is the $2\times2$ measurement error covariance matrix, and 
\begin{equation*}
	\bm{Z} = 
	\begin{pmatrix}
		1 & 0 & 0 & 0\\
    	0 & 0 & 1 & 0
	\end{pmatrix}.
\end{equation*}
That is, the observed location $\tilde{\bm{z}}_i$ is the sum of the true location $\bm{z}_i = (x_i,y_i)$ and an error term $\bm\varepsilon_i$. Using block matrix notation, the continuous state equation is
\begin{equation*}
	\bm{\omega}_{i+1} =
    \begin{pmatrix}
    	\bm{T}_i & \bm{0} \\
        \bm{0} & \bm{T}_i
    \end{pmatrix}
    \bm{\omega}_i + \bm\eta_i,\qquad
    \eta_i \sim N \left[ \bm{0}, 
    \begin{pmatrix}
    	\bm{Q}_i & \bm{0}\\
        \bm{0} & \bm{Q}_i
    \end{pmatrix}
    \right]
\end{equation*}
where
\begin{equation*}
	\bm{T}_i =
    \begin{pmatrix}
    	1 & (1-e^{-\beta_{s_i}\Delta_i})/\beta_{s_i} \\
        0 & e^{-\beta_{s_i}\Delta_i}
    \end{pmatrix}, \quad
    \bm{Q}_i =
    \begin{pmatrix}
		\text{Var}(\xi_{ci}) & \text{Cov}(\xi_{ci}, \zeta_{ci})\\
    	\text{Cov}(\xi_{ci}, \zeta_{ci}) & \text{Var}(\zeta_{ci}) 
	\end{pmatrix}.
\end{equation*}

The variances are given in Equation \ref{eqn:zeta} and \ref{eqn:xi}. The derivation of the covariance is given in Appendix S1, and yields
\begin{equation*}
	\text{Cov}(\xi_{ci}, \zeta_{ci}) = \dfrac{\sigma_{s_i}^2}
    	{2\beta_{s_i}^2} \left( 1 - 2e^{-\beta_{s_i}\Delta_i} + e^{-2\beta_{s_i}\Delta_i} \right).
\end{equation*}

Under this state-space model formulation, the Kalman filter can be used to obtain the log-likelihood of observed locations $\{ \tilde{\bm{z}}_1, \tilde{\bm{z}}_2, \dots, \tilde{\bm{z}}_n \}$. In addition, the Kalman filter and smoother provide an estimate $\hat{\bm{\omega}}_i$ of the continuous state at each time step, as well as the covariance matrix $\hat{\bm{\Sigma}}_i$ of the estimate, conditional on all observations \citep{johnson2008}. Appendix S2 gives the Kalman filter and smoother equations for the model, and the expression of the log-likelihood. For more details on state-space models and the Kalman filter, see e.g.\ \cite{durbin2012time}.

\subsection{MCMC algorithm}
\label{sec:mcmc}
We use a Markov chain Monte Carlo (MCMC) algorithm to carry out inference on the parameters of the multistate CTCRW model introduced in Section \ref{sec:multistate}, following the Metropolis-within-Gibbs approach introduced by \cite{blackwell2003}. It relies on the successive updates of the three components of the model: the reconstructed behavioural state process, the movement parameters, and the transition rates. We denote $p(\tilde{\bm{z}} \vert \bm\theta, \mathcal{S})$ the likelihood of a sequence of observed locations $\tilde{\bm{z}}$, given the movement parameters $\bm\theta$ and the reconstructed state sequence $\mathcal{S}$, as obtained from the Kalman filter presented in Section \ref{sec:kalman}.

We initialise the state sequence to $\mathcal{S}^{(0)}$, and the movement parameters to $\bm{\theta}^{(0)}$. Then, for $K$ iterations ($k = 1, \dots, K$), we run the three following steps alternately.

\paragraph{Update of the behavioural state process}
The evaluation of the likelihood of the model, described in Section \ref{sec:kalman}, is conditional on the sequence of underlying states. In practice, the states are generally not observed, such that we need to impute them. The sequence of states $\mathcal{S}$ is composed of the times of the state transitions, and the values of the states. At each iteration, an updated state sequence $\mathcal{S}^*$ is proposed as follows. We choose an interval $[t_a,t_b]$, where $a$ and $b$ are two integers such that $t_1 \leq t_a < t_b \leq t_n$. We simulate the state process $s_t$ from $t_a$ to $t_b$, conditional on $s_{t_a}$ and $s_{t_b}$, e.g.\ using the endpoint-conditioned continuous-time Markov process simulation methods from \cite{hobolth2009}. The proposed sequence of states $\mathcal{S}^*$ remains identical to $\mathcal{S}^{(k-1)}$ outside $[t_a,t_b]$. The acceptance ratio for $\mathcal{S}^*$ is
\begin{equation*}
	r = \dfrac{p(\tilde{\bm{z}} \vert \bm{\theta}^{(k-1)}, \mathcal{S}^*)}
    	{p(\tilde{\bm{z}} \vert \bm{\theta}^{(k-1)}, \mathcal{S}^{(k-1)})}.
\end{equation*}
The proposed state process reconstruction is accepted with probability $\min(1,r)$. The length of the interval $[t_a,t_b]$ over which the state sequence is updated is a tuning parameter of the sampler, and updates over longer intervals are generally less likely to be accepted.

\paragraph{Update of the movement parameters}
Denote $\bm\theta$ the vector of parameters of the movement process, i.e.\ $\bm\theta = (\beta_1, \dots, \beta_N, \sigma_1, \dots, \sigma_N)$ for a $N$-state model. We use a Metropolis-Hastings step to update the movement parameters. At iteration $k$, we propose new movement parameters $\bm{\theta}^*$, from a proposal density $q(\bm{\theta}^* \vert \bm{\theta}^{(k-1)})$, and the acceptance ratio is
\begin{equation*}
	r = \dfrac{p(\tilde{\bm{z}} \vert \bm{\theta}^*, \mathcal{S}^{(k)}) q(\bm{\theta}^{(k-1)} \vert \bm{\theta}^*)}
	{p(\tilde{\bm{z}} \vert \bm{\theta}, \mathcal{S}^{(k)}) q(\bm{\theta}^* \vert \bm{\theta}^{(k-1)})}.
\end{equation*}
The parameters are updated to $\bm{\theta}^*$ with probability $\min(1,r)$. Note that, if the proposal distribution is symmetric such that $\forall \bm{\theta}_1, \bm{\theta}_2, q(\bm{\theta}_1 \vert \bm{\theta}_2) = q(\bm{\theta}_2 \vert \bm{\theta}_1)$, then $r$ simplifies to
\begin{equation*}
	r = \dfrac{p(\tilde{\bm{z}} \vert \bm{\theta}^*, \mathcal{S}^{(k)})}
    	{p(\tilde{\bm{z}} \vert \bm{\theta}^{(k-1)}, \mathcal{S}^{(k)})}.
\end{equation*}
In practice, a standard choice is to use a multivariate normal proposal distribution on the working scale of the parameters (in this case, on the log scale). Its variance can be tuned to obtain different acceptance rates, and covariance structure can be added to explore the parameter space more efficiently.

\paragraph{Update of the transition rates}
Following \cite{blackwell2003}, using conjugate priors, the transition rates can be directly sampled from their posterior distribution, which is known conditionally on the reconstructed state sequence $\mathcal{S}$. We find it convenient to parametrise the generator matrix as
\begin{equation*}
	\bm{\Lambda} = 
    \begin{pmatrix}
		-\lambda_1 & \lambda_1 p_{12} &  \cdots & \lambda_1 p_{1N} \\
        \lambda_2 p_{21} & -\lambda_2 &  \cdots & \lambda_2 p_{2N} \\
        \vdots & \vdots & \ddots & \vdots \\
        \lambda_N p_{N1} & \lambda_N p_{N2} & \cdots & -\lambda_N \\
	\end{pmatrix}
\end{equation*}
where $\lambda_i>0$ is the rate of transition out of state $i$, and the $p_{ij} \in [0,1]$ are the transition probabilities out of state $i$. For each state $i$, they satisfy $\sum_j p_{ij} = 1$. The transition rates and the transition probabilities can be sampled separately.

For each $i \in \{1,\dots,N\}$, we denote $n_i$ the number of time intervals spent in state $i$, and $(\tau_i^{(1)}, \tau_i^{(2)}, \dots, \tau_i^{(n_i)})$ their lengths. These dwell times are exponentially distributed with rate $\lambda_i$. The conjugate prior of the exponential distribution is the gamma distribution such that, with the prior
\begin{equation*}
	\lambda_i \sim \text{gamma}(\alpha_1,\alpha_2),
\end{equation*}
the transition rates are sampled from the posterior distribution
\begin{equation*}
	\lambda_i \vert \mathcal{S} \sim \text{gamma} \left(\alpha_1 + n_i,\ \alpha_2 + \sum_{j=1}^{n_i} \tau_i^{(j)} \right). 
\end{equation*}

For $i \in \{1,\dots,N\}$ and $j \in \{1,\dots,N\}$ such that $i \neq j$, we denote $n_{ij}$ the number of transitions from state $i$ to state $j$, and $n_i = \sum_j n_{ij}$ the number of transitions out of state $i$. Then,
\begin{equation*}
	n_{i1}, n_{i2}, \dots, n_{i_N} \sim \text{multinom}(n_i, p_{i1}, p_{i2}, \dots, p_{iN}).
\end{equation*}

The conjugate prior of the multinomial distribution is the Dirichlet distribution such that, with the prior
\begin{equation*}
	p_{i1},p_{i2},\dots,p_{iN} \sim \text{Dir}(\kappa_{i1},\kappa_{i2},\dots,\kappa_{iN}),
\end{equation*}
the posterior distribution of the transition probabilities is
\begin{equation*}
	p_{i1},p_{i2},\dots,p_{iN} \vert \mathcal{S} \sim \text{Dir}(\kappa_{i1}+n_{i1}, \kappa_{i2}+n_{i2}, \dots, \kappa_{iN}+n_{iN}).
\end{equation*}

\section{Simulation study} \label{sec:simulation}
We simulated a trajectory from the multistate CTCRW model described in Section \ref{sec:multistate}, and used the MCMC algorithm described in Section \ref{sec:inference} to infer the hidden state sequence and the movement parameters from the simulated data.

\subsection{Simulated data}
We simulated 10,000 locations from a 2-state model at a fine time scale (every 0.1 time unit), and thinned them by keeping 10\% of the points at random (i.e.\ 1000 irregularly-spaced locations), to emulate real movement data. The time intervals in the resulting data set ranged between 0.1 and 8 time units. 

The movement parameters and switching rates of the simulated process were chosen as
\begin{equation*}
	(\beta_1,\beta_2) = (1,0.3),\quad (\sigma_1, \sigma_2) = (1,3),\quad \bm{\Lambda} =
	\begin{pmatrix}
  		-0.03 & 0.03 \\
  		0.03 & -0.03
  	\end{pmatrix}.
\end{equation*}

\begin{figure}[htbp]
	\centering
    \includegraphics[width=0.5\textwidth]{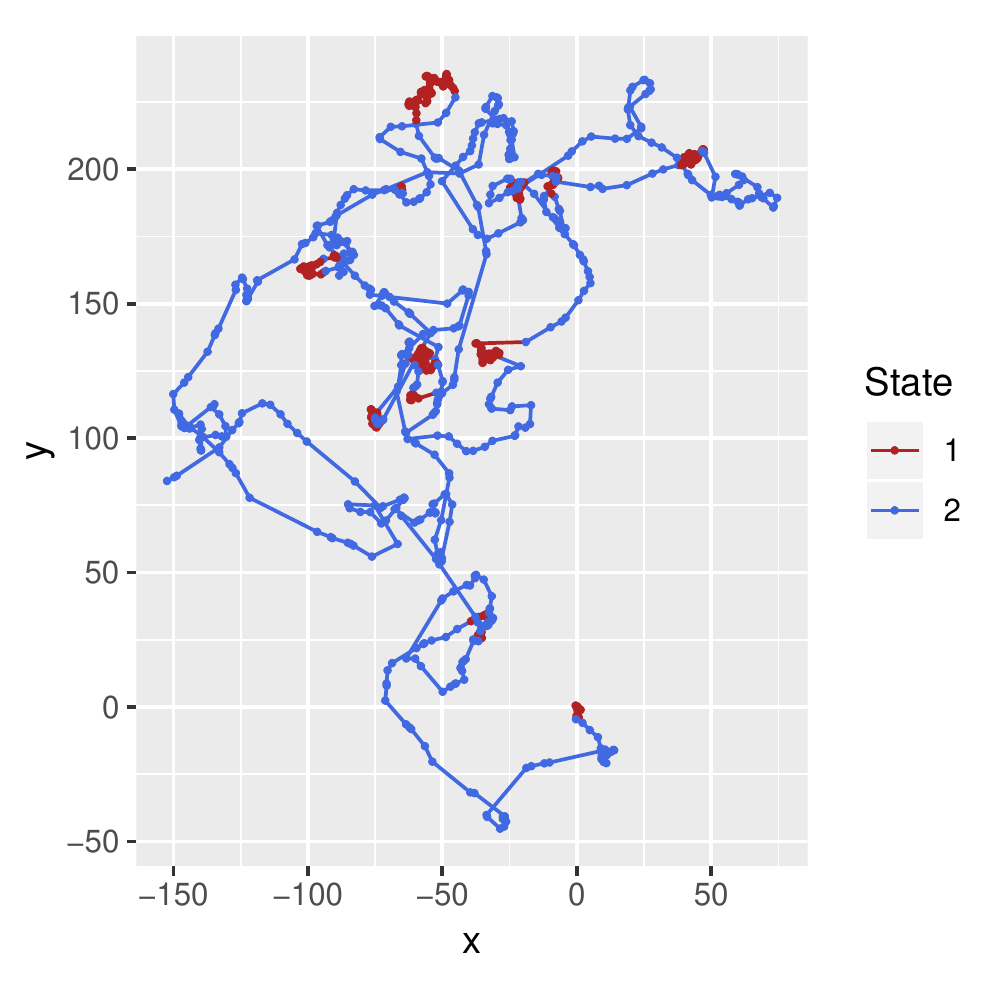}
    \caption{Track simulated from a 2-state CTCRW model.}
    \label{fig:simtrack}
\end{figure}

In state 1, the variance was smaller and the reversion to the mean larger, which resulted in slower and more sinuous movement (perhaps analogous to ``area-restricted search'' behaviour). State 2 corresponded to faster and more directed movement. The transition rates were chosen such that the process would on average stay about 30 time units in a state before switching to the other state. The simulated track (after thinning) is shown in Figure \ref{fig:simtrack}.

\subsection{Estimation}
We initialised the reconstructed state sequence by classifying each observation randomly as being in state 1 or state 2, with probability 0.5 each. At each iteration, the state process was updated over a randomly-selected interval of five time steps. We used independent normal proposal distributions (on the working log scale) to update the movement parameters. The proposal variances were tuned based on initial test runs, to obtain near-optimal acceptance rates. We chose normal prior distributions on the working scale for the movement parameters, centred on the true values of the parameters, and with large variances.

\begin{figure}[htbp]
	\centering
    \includegraphics[width=0.6\textwidth]{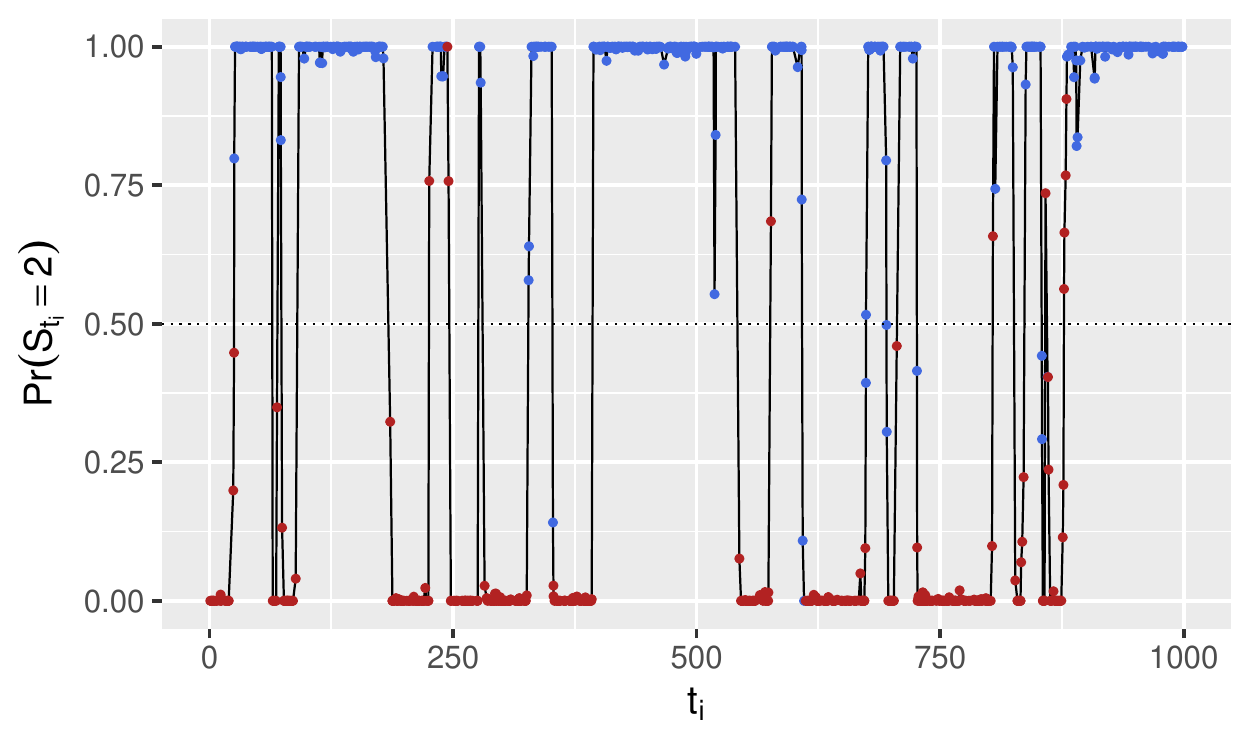}
    \caption{Posterior probabilities of being in state 2 at the times of the observations. The true (simulated) states are shown by the colours (red: state 1, blue: state 2). }
    \label{fig:stateProbs}
\end{figure}

We ran $10^5$ MCMC iterations, which took around 20 min on an 2GHz i5 CPU, and discarded the first $5 \times 10^4$ as burn-in. Figure \ref{fig:stateProbs} shows the posterior probabilities of being in state 2 at the times of the observations, to compare with the ``true'' simulated state sequence. For each $i=1,\dots,n$, we calculated the posterior probability as the proportion of reconstructions of the state process in which $\bm{x}_i$ was classified in state 2. (It would have been equivalent to consider the posterior probability of being in state 1; the two probabilities sum to 1.) We considered that a step was misclassified if the posterior probability of being in the true state was less than 0.5. The states were correctly estimated in the vast majority of time steps, with only 2\% of misclassified steps.

\begin{figure}[htbp]
	\centering
    \includegraphics[width=0.8\textwidth]{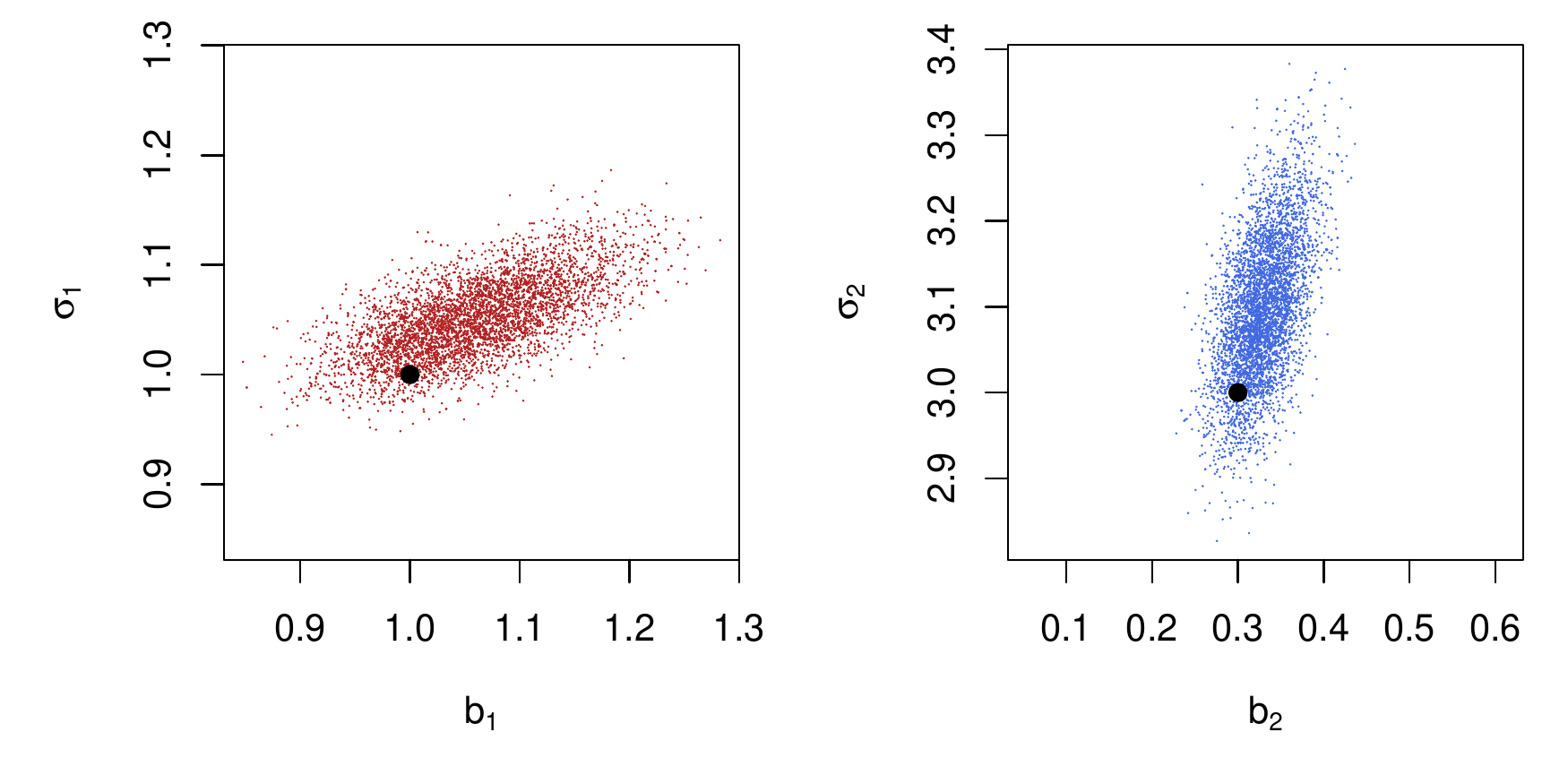}
    \caption{Posterior samples of the movement parameters in state 1 (left) and in state 2 (right), in the simulation study. The black dots are the true values of the parameters, used in the simulation. The samples are thinned to every tenth value, for visualisation purposes.}
    \label{fig:b_sigma}
\end{figure}

Figure \ref{fig:b_sigma} displays posterior samples for the state-dependent movement parameters, $\beta_1$, $\beta_2$, $\sigma_1$, and $\sigma_2$, as well as the true parameter values used in the simulation. The posterior distributions seem to appropriately estimate all movement parameters.

We were able to recover the values of the state process and of the state-dependent movement parameters from a simulated track thinned to irregular intervals. This demonstrates the ability of the method to work across temporal scales, and to cope with irregular sampling.

\section{Grey seal case study} \label{sec:greyseal}
We illustrate the use of the method described in Section \ref{sec:inference} for the analysis of a grey seal (\emph{Halichoerus grypus}) movement track. We considered a trajectory of 1875 observations, collected in the North Sea between April and August 2008, and previously described by \cite{russell2015}. The base sampling frequency was of one location every 30 minutes, but many fixes were missed, and the resulting time grid was highly irregular ($P_{0.025}$ = 27 min, $P_{0.975}$ = 8 hours). Note that the CTCRW model describes movement on a plane, and thus requires that the longitude-latitude locations be projected to UTM coordinates for the analysis.

We considered a 2-state CTCRW model, with four movement parameters to estimate: $\beta_1$, $\beta_2$, $\sigma_1$, $\sigma_2$. Similarly to the simulation study, we initialised the state reconstruction to a random sequence of 1s and 2s (with probability 0.5 each). We used independent normal proposal and prior distributions on the working scale of the movement parameters. We selected the proposal variances based on test runs, and used weakly informative prior distributions. We ran 2 million MCMC iterations, discarding the first half as burn-in. We only saved every 100th reconstructed state sequence, because of memory limitations.

\begin{figure}[htbp]
	\centering
    \includegraphics[width=0.6\textwidth]{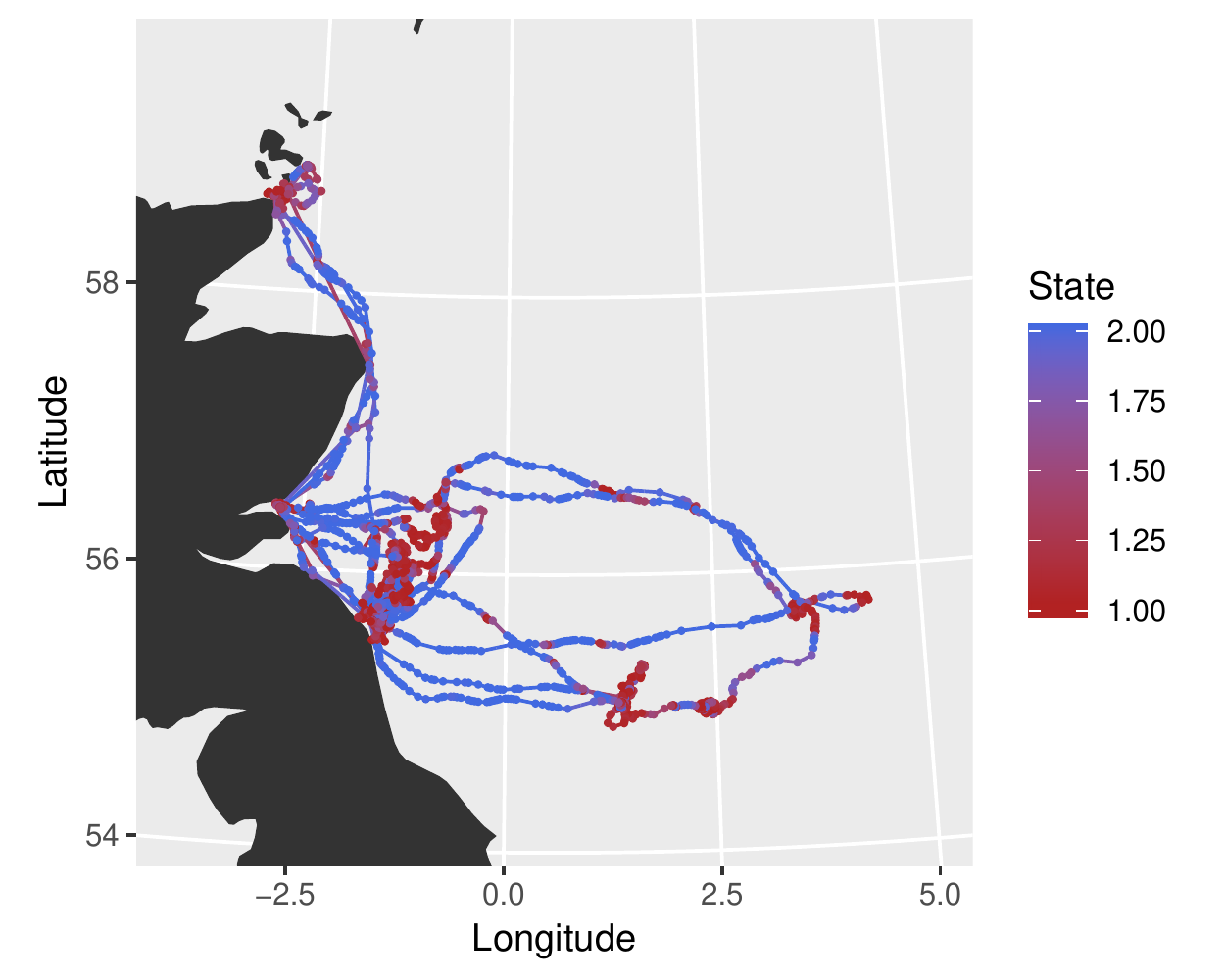}
    \caption{Grey seal track, off the East coast of Great Britain, coloured by posterior state probabilities.}
    \label{fig:sealtrack}
\end{figure}

\begin{figure}[htbp]
	\centering
    \includegraphics[width=0.8\textwidth]{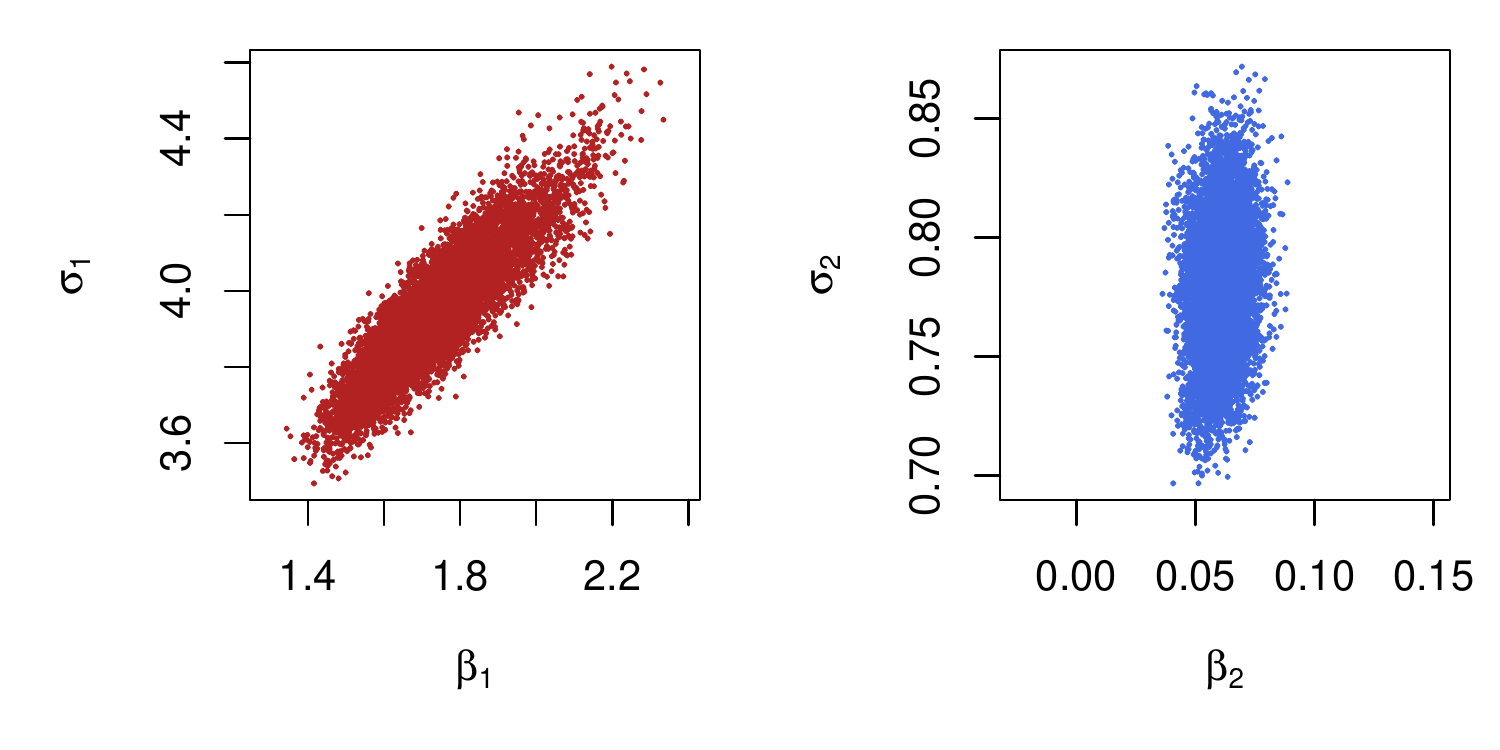}
    \caption{Posterior samples of the movement parameters in state 1 (left) and in state 2 (right), in the grey seal case study. The samples are thinned to every 100th value, for visualisation purposes.}
    \label{fig:b_sigma_seal}
\end{figure}

Figure \ref{fig:sealtrack} shows a map of the track, coloured by posterior state probabilities, and Figure \ref{fig:b_sigma_seal} shows posterior samples for the four movement parameters ($\beta_1, \beta_2, \sigma_1, \sigma_2$). State 2 captured very directed movements, corresponding to periods of transit between areas of interest, and state 1 captured more tortuous phases of the track. This can be seen in Figure \ref{fig:b_sigma_seal}: the posterior distribution of $\beta_1$ covers much larger values than that of $\beta_2$ (posterior means of 1.73 and 0.06, respectively), indicating stronger reversion to the mean in state 1, and thus less movement persistence. There were no signs of label switching in the posterior samples; if there were, a straightforward solution would be to constrain $(\beta_1,\beta_2)$ and $(\sigma_1,\sigma_2)$ to be ordered \citep{mcclintock2014}.

\begin{figure}[htbp]
	\centering
    \includegraphics[width=0.7\textwidth]{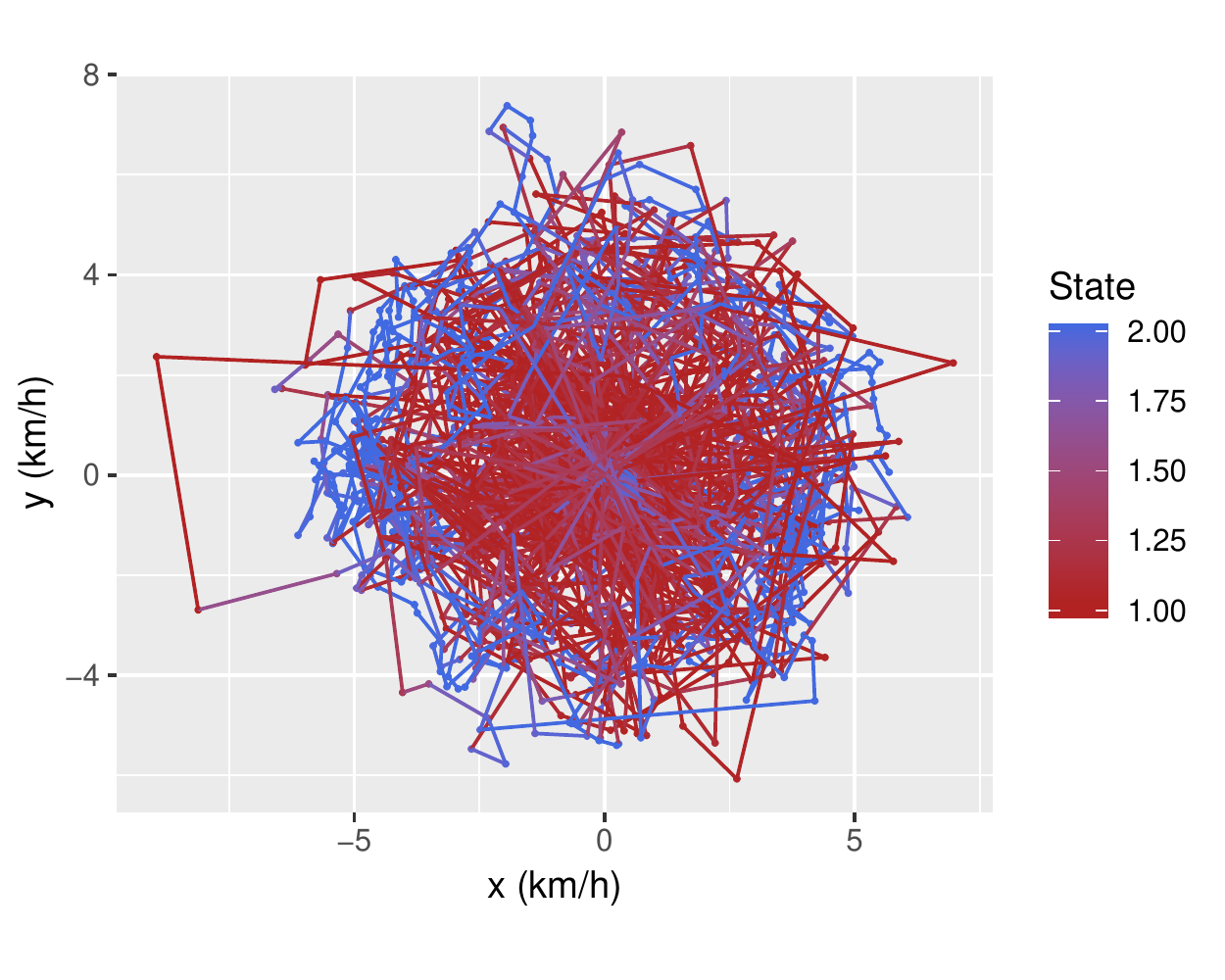}
    \caption{Predicted velocities from the grey seal example, obtained with the mean posterior movement parameter estimates. The colour reflects the mean (over the sampled state sequences) of the states to which each observation was allocated.}
    \label{fig:velocity_seal}
\end{figure}

The Kalman filter and smoother recursions given in Appendix S2 can be used to compute estimated velocities at the times of the observations. The velocities obtained with the mean posterior movement parameters are displayed in Figure \ref{fig:velocity_seal}, and coloured by posterior state probabilities. The strong movement autocorrelation in state 2 can be seen in the outer rim of the plot, where the velocity sometimes persists with little variation over many time steps. The Kalman algorithm can also provide estimates of the locations (and possibly velocities) of the animal -- and associated standard errors -- on any time grid, e.g.\ on a finer time grid than that of the observations.

An alternative parametrisation of the CTCRW is defined with $\tau = 1/\beta$ and $\nu = \sqrt{\pi} \sigma/(2 \sqrt{\beta})$ \citep{gurarie2017}. The parameter $\tau$ is the time interval over which the autocorrelation function of the velocity process decreases by a factor $e$, such that larger values indicate longer-term movement persistence. The parameter $\nu$ is the mean speed of movement of the animal. We transformed the posterior samples of movement parameters to obtain estimates of $\tau$ and $\nu$ in both states. In the following, we report posterior mean estimates, and histograms of the posterior samples for the transformed parameters can be found in Appendix S3. The posterior means were $\hat\tau_1 = 0.55$h, $\hat\tau_2 = 17.21$h, $\hat\nu_1 = 2.63$km.h$^{-1}$, and $\hat\nu_2 = 2.90$km.h$^{-1}$. This indicates that the two states are very similar in terms of the speed of movement, but that the autocorrelation function of the velocity drops much faster in state 1 than in state 2.

The posterior samples of the transition rates can be used to derive mean dwell times in each state, and long-term activity budgets. The dwell times in state $i$ follow an exponential distribution with rate $\lambda_i$ (the rate of transition out of state $i$). The mean dwell time can thus be derived as $d_i = 1/\lambda_i$. In the grey seal analysis, the posterior means for the mean dwell time were $\hat{d}_1 = 7.5$h in state 1, and $\hat{d}_2 = 7.8$h in state 2, indicating similar dwell times in both states. Activity budgets refer to the proportion of time spent by an animal in each of its behavioural states \citep{russell2015}. In a time-homogeneous state-switching model, an estimate of the long-term activity budget can be calculated as the stationary distribution of the underlying Markov process. The stationary distribution of a $N$-state Markov process is the vector $\bm\pi = (\pi_1, \dots, \pi_N)$ which satisfies $\bm\pi \bm\Lambda = \bm{0}$, subject to the constraint $\sum_{i=1}^N \pi_i = 1$, where $\bm\Lambda$ is the generator matrix defined in Equation \ref{eqn:generator}. In the 2-state case, solving the equation yields $\pi_1 = \lambda_2/(\lambda_1+\lambda_2)$ and $\pi_2 = \lambda_1/(\lambda_1+\lambda_2)$. The posterior mean estimate for the stationary distribution was $(\hat\pi_1, \hat\pi_2) = (0.48,0.52)$, i.e.\ the seal will tend to spend roughly the same proportion of time in both states, in the long term. Histograms of posterior draws for the dwell times and stationary distribution are displayed in Appendix S3.

\section{Discussion}
We presented a Bayesian framework to infer discrete behavioural states and movement parameters from a multistate continuous-time model of animal movement. The continuous-time formulation can accommodate irregular time intervals, and is consistent across temporal scales. The conditional likelihood of the model, used in the MCMC algorithm, is implemented using the Kalman filter, making it relatively fast and allowing for the inclusion of measurement error.

The inferential approach introduced in this paper could in principle be used to implement a state-switching version of the OUF model described by \cite{fleming2014}. The OUF process is a generalisation of the CTCRW used in this paper, and of the Ornstein-Uhlenbeck location process used e.g.\ by \cite{blackwell2016}. It features both persistence in velocity and long-term attraction towards a point in space, making it a very flexible model of animal movement. Like the CTCRW, it can be written as a state-space model, and the Kalman filter can be used to derive the likelihood of the model \citep{fleming2017}. The MCMC algorithm described in Section \ref{sec:mcmc} could then be used to fit a multistate OUF model to animal movement. However, the OUF process has five parameters (against 2 only for the CTCRW), which could make estimation more challenging. More generally, this methodology could be applied to a model switching between processes with different formulations (e.g.\ a 2-state model switching between a CTCRW and OUF process). 

Analyses of animal movement and behaviour often combine telemetry and environmental data. In state-switching models, the effect of environmental (or other) covariates on the transition probabilities is of particular interest, and is used to uncover the drivers of animal behavioural and movement decisions \citep{patterson2009, bestley2012, blackwell2016}. The MCMC algorithm of Section \ref{sec:mcmc} could be extended, following \cite{blackwell2016}, to allow for the inclusion of covariates in the state-switching dynamics.

\subsection*{Acknowledgements}
TM was supported by the Centre for Advanced Biological Modelling at the University of Sheffield, funded by the Leverhulme Trust, award number DS-2014-081. We thank Debbie Russell and Esther Jones for helpful advice about the grey seal data set. The grey seal telemetry data were provided by the Sea Mammal Research Unit; their collection was supported by funding from the Natural Environment Research Council to SMRU.

\bibliographystyle{apalike}
\bibliography{refs.bib}

\begin{thebibliography}{}

\bibitem[Berman, 1994]{berman1994}
Berman, S.~M. (1994).
\newblock A bivariate markov process with diffusion and discrete components.
\newblock {\em Stochastic Models}, 10(2):271--308.

\bibitem[Bestley et~al., 2012]{bestley2012}
Bestley, S., Jonsen, I.~D., Hindell, M.~A., Guinet, C., and Charrassin, J.-B.
  (2012).
\newblock Integrative modelling of animal movement: incorporating in situ
  habitat and behavioural information for a migratory marine predator.
\newblock {\em Proceedings of the Royal Society of London B: Biological
  Sciences}, page rspb20122262.

\bibitem[Blackwell, 1997]{blackwell1997}
Blackwell, P. (1997).
\newblock Random diffusion models for animal movement.
\newblock {\em Ecological Modelling}, 100(1-3):87--102.

\bibitem[Blackwell, 2003]{blackwell2003}
Blackwell, P. (2003).
\newblock Bayesian inference for markov processes with diffusion and discrete
  components.
\newblock {\em Biometrika}, 90(3):613--627.

\bibitem[Blackwell et~al., 2016]{blackwell2016}
Blackwell, P.~G., Niu, M., Lambert, M.~S., and LaPoint, S.~D. (2016).
\newblock Exact bayesian inference for animal movement in continuous time.
\newblock {\em Methods in Ecology and Evolution}, 7(2):184--195.

\bibitem[Codling and Hill, 2005]{codling2005}
Codling, E. and Hill, N. (2005).
\newblock Sampling rate effects on measurements of correlated and biased random
  walks.
\newblock {\em Journal of Theoretical Biology}, 233(4):573--588.

\bibitem[Dunn and Gipson, 1977]{dunn1977}
Dunn, J.~E. and Gipson, P.~S. (1977).
\newblock Analysis of radio telemetry data in studies of home range.
\newblock {\em Biometrics}, pages 85--101.

\bibitem[Durbin and Koopman, 2012]{durbin2012time}
Durbin, J. and Koopman, S.~J. (2012).
\newblock {\em Time series analysis by state space methods}, volume~38.
\newblock OUP Oxford.

\bibitem[Fleming et~al., 2014]{fleming2014}
Fleming, C.~H., Calabrese, J.~M., Mueller, T., Olson, K.~A., Leimgruber, P.,
  and Fagan, W.~F. (2014).
\newblock From fine-scale foraging to home ranges: a semivariance approach to
  identifying movement modes across spatiotemporal scales.
\newblock {\em The American Naturalist}, 183(5):E154--E167.

\bibitem[Fleming et~al., 2017]{fleming2017}
Fleming, C.~H., Sheldon, D., Gurarie, E., Fagan, W.~F., LaPoint, S., and
  Calabrese, J.~M. (2017).
\newblock K{\'a}lm{\'a}n filters for continuous-time movement models.
\newblock {\em Ecological Informatics}, 40:8--21.

\bibitem[Gurarie et~al., 2017]{gurarie2017}
Gurarie, E., Fleming, C.~H., Fagan, W.~F., Laidre, K.~L., Hern{\'a}ndez-Pliego,
  J., and Ovaskainen, O. (2017).
\newblock Correlated velocity models as a fundamental unit of animal movement:
  synthesis and applications.
\newblock {\em Movement Ecology}, 5(1):13.

\bibitem[Hobolth and Stone, 2009]{hobolth2009}
Hobolth, A. and Stone, E.~A. (2009).
\newblock Simulation from endpoint-conditioned, continuous-time markov chains
  on a finite state space, with applications to molecular evolution.
\newblock {\em The annals of applied statistics}, 3(3):1204.

\bibitem[Hooten et~al., 2017]{hooten2017}
Hooten, M.~B., Johnson, D.~S., McClintock, B.~T., and Morales, J.~M. (2017).
\newblock {\em Animal movement: statistical models for telemetry data}.
\newblock CRC Press.

\bibitem[Horne et~al., 2007]{horne2007}
Horne, J.~S., Garton, E.~O., Krone, S.~M., and Lewis, J.~S. (2007).
\newblock Analyzing animal movements using brownian bridges.
\newblock {\em Ecology}, 88(9):2354--2363.

\bibitem[Johnson, 2017]{johnson2017}
Johnson, D.~S. (2017).
\newblock {\em crawl: Fit Continuous-Time Correlated Random Walk Models to
  Animal Movement Data}.
\newblock R package version 2.1.1.

\bibitem[Johnson et~al., 2008]{johnson2008}
Johnson, D.~S., London, J.~M., Lea, M.-A., and Durban, J.~W. (2008).
\newblock Continuous-time correlated random walk model for animal telemetry
  data.
\newblock {\em Ecology}, 89(5):1208--1215.

\bibitem[Jonsen et~al., 2005]{jonsen2005}
Jonsen, I.~D., Flemming, J.~M., and Myers, R.~A. (2005).
\newblock Robust state--space modeling of animal movement data.
\newblock {\em Ecology}, 86(11):2874--2880.

\bibitem[Langrock et~al., 2012]{langrock2012}
Langrock, R., King, R., Matthiopoulos, J., Thomas, L., Fortin, D., and Morales,
  J.~M. (2012).
\newblock Flexible and practical modeling of animal telemetry data: hidden
  markov models and extensions.
\newblock {\em Ecology}, 93(11):2336--2342.

\bibitem[McClintock, 2017]{mcclintock2017}
McClintock, B.~T. (2017).
\newblock Incorporating telemetry error into hidden markov models of animal
  movement using multiple imputation.
\newblock {\em Journal of Agricultural, Biological and Environmental
  Statistics}, 22(3):249--269.

\bibitem[McClintock et~al., 2014]{mcclintock2014}
McClintock, B.~T., Johnson, D.~S., Hooten, M.~B., Ver~Hoef, J.~M., and Morales,
  J.~M. (2014).
\newblock When to be discrete: the importance of time formulation in
  understanding animal movement.
\newblock {\em Movement ecology}, 2(1):21.

\bibitem[Morales et~al., 2004]{morales2004}
Morales, J.~M., Haydon, D.~T., Frair, J., Holsinger, K.~E., and Fryxell, J.~M.
  (2004).
\newblock Extracting more out of relocation data: building movement models as
  mixtures of random walks.
\newblock {\em Ecology}, 85(9):2436--2445.

\bibitem[Parton and Blackwell, 2017]{parton2017}
Parton, A. and Blackwell, P.~G. (2017).
\newblock Bayesian inference for multistate `step and turn' animal movement in
  continuous time.
\newblock {\em Journal of Agricultural, Biological and Environmental
  Statistics}, 22(3):373--392.

\bibitem[Patterson et~al., 2009]{patterson2009}
Patterson, T.~A., Basson, M., Bravington, M.~V., and Gunn, J.~S. (2009).
\newblock Classifying movement behaviour in relation to environmental
  conditions using hidden markov models.
\newblock {\em Journal of Animal Ecology}, 78(6):1113--1123.

\bibitem[Patterson et~al., 2017]{patterson2017}
Patterson, T.~A., Parton, A., Langrock, R., Blackwell, P.~G., Thomas, L., and
  King, R. (2017).
\newblock Statistical modelling of individual animal movement: an overview of
  key methods and a discussion of practical challenges.
\newblock {\em AStA Advances in Statistical Analysis}, 101(4):399--438.

\bibitem[Preisler et~al., 2013]{preisler2013}
Preisler, H.~K., Ager, A.~A., and Wisdom, M.~J. (2013).
\newblock Analyzing animal movement patterns using potential functions.
\newblock {\em Ecosphere}, 4(3):1--13.

\bibitem[Russell et~al., 2015]{russell2015}
Russell, D.~J., McClintock, B.~T., Matthiopoulos, J., Thompson, P.~M.,
  Thompson, D., Hammond, P.~S., Jones, E.~L., MacKenzie, M.~L., Moss, S., and
  McConnell, B.~J. (2015).
\newblock Intrinsic and extrinsic drivers of activity budgets in sympatric grey
  and harbour seals.
\newblock {\em Oikos}, 124(11):1462--1472.

\bibitem[Schl{\"a}gel and Lewis, 2016]{schlagel2016}
Schl{\"a}gel, U.~E. and Lewis, M.~A. (2016).
\newblock Robustness of movement models: can models bridge the gap between
  temporal scales of data sets and behavioural processes?
\newblock {\em Journal of mathematical biology}, 73(6-7):1691--1726.

\bibitem[Uhlenbeck and Ornstein, 1930]{uhlenbeck1930}
Uhlenbeck, G.~E. and Ornstein, L.~S. (1930).
\newblock On the theory of the brownian motion.
\newblock {\em Physical review}, 36(5):823.

\end{thebibliography}

\setcounter{equation}{0}
\renewcommand\thefigure{S\arabic{figure}}
\setcounter{figure}{0}  
\newpage

\section*{Appendix S1. CTCRW transition density}
The CTCRW, or integrated Ornstein-Uhlenbeck process $(\bm{z}_t)_{t>0}$, is defined as the solution to
\begin{equation*}
\begin{cases}
	d\bm{z}_t = \bm{v}_t dt,\\
    d\bm{v}_t = -\beta \bm{v}_t dt + \sigma d\bm{w}_t.
\end{cases}
\end{equation*}

We consider here the isotropic case only, so it is sufficient to solve this system of equations in either dimension. In the following, we consider the univariate processes $z_t$, $v_t$, and $w_t$.

\subsection*{Solution to the Ornstein-Uhlenbeck equation}
We first derive the solution for the equation of the velocity $v_t$,
\begin{equation}
	\label{eq1}
	dv_t = -\beta v_t dt + \sigma dw_t
\end{equation}

Note that
\begin{equation}
  \label{eq2}
  d(e^{\beta t} v_t) = \beta e^{\beta t} v_t + e^{\beta t} dv_t
\end{equation}

Then, from (\ref{eq1}) and (\ref{eq2}),
\begin{equation*}
  d(e^{\beta t} v_t) = e^{\beta t} \sigma dw_t
\end{equation*}

Integrating both sides between $t$ and $t+\delta$,
\begin{equation*}
  e^{\beta(t+\delta)} v_{t+\delta} - e^{\beta t} v_t = \sigma \int_{s=t}^{t+\delta} e^{\beta s} dw_s,
\end{equation*}
and we obtain the solution
\begin{equation}
\label{eqn:OUsolution}
  v_{t+\delta} = e^{-\beta\delta} v_t + \sigma \int_{s=t}^{t+\delta} e^{-\beta(t+\delta-s)} dw_s.
\end{equation}

Denote $\zeta(\delta) = \sigma \int_{s=t}^{t+\delta} e^{-\beta(t+\delta-s)} dw_s$. By property of the It\^o integral, $\zeta(\delta)$ is a Gaussian random variable with mean $0$ and variance
\begin{equation*}
  \text{Var}(\zeta(\delta)) = E \left[\zeta(\delta)^2 \right]
  = \sigma^2 E \left[ \left( \int_{s=t}^{t+\delta} e^{-\beta(t+\delta-s)} dw_s \right)^2 \right]
\end{equation*}

Using the It\^o isometry,
\begin{align*}
  \text{Var}(\zeta(\delta)) & = \sigma^2 \int_{s=t}^{t+\delta} e^{-2\beta(t+\delta-s)} ds\\
  & = \sigma^2 \left[ \dfrac{e^{-2\beta(t+\delta-s)}}{2\beta} \right]_{s=t}^{t+\delta} \\
  & = \dfrac{\sigma^2}{2\beta} (1-e^{-2\beta\delta}) 
\end{align*}

Eventually,
\begin{equation*}
	\zeta (\delta) \sim N \left( 0,\  \dfrac{\sigma^2}{2\beta} (1-e^{-2\beta\delta})  \right)
\end{equation*}

\subsection*{Solution to the integrated Ornstein-Uhlenbeck equation}
We can now solve for the location process $z_t$. Integrating both sides of $dz_t = v_t dt$ between $t$ and $t+\delta$, and using the solution found in Equation \ref{eqn:OUsolution}, we have
\begin{equation*}
	z_{t+\delta} - z_t = 
    \int_{s=t}^{t+\delta} v_s ds = 
    \int_{s=t}^{t+\delta} \left( e^{-\beta(s-t)} v_t + \sigma \int_{u=t}^s e^{-\beta(s-u)} dw_u \right) ds
\end{equation*}

Thus,
\begin{align*}
	z_{t+\delta} & = z_t + v_t \int_{s=t}^{t+\delta} e^{-\beta(s-t)} ds + 
    	\sigma \int_{s=t}^{t+\delta} \int_{u=t}^s e^{-\beta(s-u)} dw_u ds \\
    & = z_t + v_t \left[ -\dfrac{e^{-\beta(s-t)}}{\beta} \right]_{s=t}^{t+\delta} + 
    	\sigma \int_{u=t}^{t+\delta} \int_{s=u}^{t+\delta} e^{-\beta(s-u)} ds dw_s \\
    & = z_t + \left( \dfrac{1-e^{-\beta\delta}}{\beta} \right) v_t + \sigma \int_{u=t}^{t+\delta} 
    	\left[ -\dfrac{e^{-\beta(s-u)}}{\beta} \right]_{s=u}^{t+\delta} dw_u \\
    & = z_t + \left( \dfrac{1-e^{-\beta\delta}}{\beta} \right) v_t + \dfrac{\sigma}{\beta} \int_{u=t}^{t+\delta} 
    	(1 - e^{-\beta(t+\delta-u)}) dw_u \\
\end{align*}

The location process $z_t$ also has a normal transition density, with mean
\begin{equation*}
	E(z_{t+\delta} \vert z_t, v_t) = z_t + \left( \dfrac{1-e^{-\beta\delta}}{\beta} \right) v_t.
\end{equation*}

Denote $\xi(\delta)$ the Gaussian error term
\begin{equation*}
	\xi(\delta) = \dfrac{\sigma}{\beta} \int_{u=t}^{t+\delta} (1+e^{-\beta(t+\delta-u)}) dw_u.
\end{equation*}

The variance of the transition density of the location process is given by $\text{Var}(\delta)$,
\begin{equation*}
	\text{Var}(\xi(\delta)) = \left( \dfrac{\sigma}{\beta} \right)^2 
    	E \left[ \left( \int_{u=t}^{t+\delta} (1-e^{-\beta(t+\delta-u)}) dw_u \right)^2 \right].
\end{equation*}

Using the It\^o isometry,
\begin{align*}
	\text{Var}(\xi(\delta)) & = \left( \dfrac{\sigma}{\beta} \right)^2 
    	\int_{u=t}^{t+\delta} (1-e^{-\beta(t+\delta-u)})^2 du \\
    & = \left( \dfrac{\sigma}{\beta} \right)^2 
    	\int_{u=t}^{t+\delta} (1 + e^{-2\beta(t+\delta-u)} - 2e^{-\beta(t+\delta-u)}) du \\
    & = \left( \dfrac{\sigma}{\beta} \right)^2 
    	\left[ u + \dfrac{e^{-2\beta(t+\delta-u)}}{2\beta} - \dfrac{2e^{-\beta(t+\delta-u)}}{\beta}
        \right]_{u=t}^{t+\delta} \\
    & = \left( \dfrac{\sigma}{\beta} \right)^2 
    	\left( \delta + \dfrac{1 - e^{-2\beta\delta}}{2\beta} - \dfrac{2 (1-e^{-\beta\delta})}{\beta} \right) \\
\end{align*}

Finally, we obtain
\begin{equation*}
	\xi(\delta) \sim N \left( 0,\ \left( \dfrac{\sigma}{\beta} \right)^2 
    	\left( \delta + \dfrac{1 - e^{-2\beta\delta}}{2\beta} - \dfrac{2 (1-e^{-\beta\delta})}{\beta} \right) \right)
\end{equation*}

\subsection*{Covariance of the location process and velocity process}
To formulate the CTCRW as a state-space model, we also need the covariance of the location process and the velocity process, i.e.\ $\text{Cov}(\xi(\delta),\zeta(\delta))$.

\begin{align*}
	\text{Cov}(\zeta(\delta),\xi(\delta)) & = E(\zeta(\delta)\xi(\delta)) \\
    & = E \left[ \left( \sigma \int_{s=t}^{t+\delta} e^{-\beta(t+\delta-s)} dw_s \right) 
    	\left( \dfrac{\sigma}{\beta} \int_{s=t}^{t+\delta} (1-e^{-\beta(t+\delta-s)}) dw_s \right) \right].
\end{align*}

Using the It\^o isometry,
\begin{align*}
	\text{Cov}(\zeta(\delta),\xi(\delta)) & = 
    	\dfrac{\sigma^2}{\beta} \int_{s=t}^{t+\delta} e^{-\beta(t+\delta-s)} (1-e^{-\beta(t+\delta-s)}) ds \\
    & = \dfrac{\sigma^2}{\beta} \int_{s=t}^{t+\delta} (e^{-\beta(t+\delta-s)} - e^{-2\beta(t+\delta-s)} ds \\
    & = \dfrac{\sigma^2}{\beta} \left[ \dfrac{e^{-\beta(t+\delta-s)}}{\beta} - 
    	\dfrac{e^{-2\beta(t+\delta-s)}}{2\beta} \right]_{s=t}^{t+\delta} \\
    & = \dfrac{\sigma^2}{\beta} \left( \dfrac{1-e^{-2\beta\delta}}{\beta} - 
    	\dfrac{1-e^{-2\beta\delta}}{2\beta} \right) \\
    & = \dfrac{\sigma^2}{2\beta^2} \left( 1 - 2e^{-\beta\delta} + e^{-2\beta\delta} \right)
\end{align*}

\newpage
\section*{Appendix S2. Kalman filtering and smoothing}
Here we give the equations of the Kalman filter and smoother, for the multistate CTCRW model. These equations can be found for the single-state CTCRW in Appendix B of \cite{johnson2008}, and in the case of a general state-space model in \cite{durbin2012time}.

\paragraph{Notation} We use the following notation:
\begin{itemize}
\item $\{ \tilde{\bm{z}}_1 , \dots, \tilde{\bm{z}}_n \}$ are the observed locations, and $\{ t_1, \dots, t_n \}$ the corresponding times. Note that $\tilde{\bm{z}}_i$ is treated as missing data if $t_i$ corresponds to a behavioural transition (see below for processing of missing data).
\item $\tilde{\bm{z}}_i = (\tilde{x}_i, \tilde{y}_i)^\top$.
\item $s_i$ is the behavioural state between $t_i$ and $t_{i+1}$.
\end{itemize}

\paragraph{Filtering} A standard choice to initialise the state estimate is $\hat{\bm{a}}_1 = (\tilde{x}_1, 0, \tilde{y}_1, 0)^\top$, i.e.\ the mean initial location is the first observation $\tilde{\bm{z}}_1$, and the mean initial velocity is $\bm{0}$. The estimate covariance matrix $\hat{\bm{P}}_1$ is typically taken to be diagonal, and its elements measure the uncertainty on the initial state estimate $\hat{\bm{a}}_1$. 

Then, for $i = 1, 2, \dots, n-1$, the Kalman filter equations are
\begin{itemize}
\item $\bm{u}_i = \tilde{\bm{z}}_i - \bm{Z} \hat{\bm{a}}_i$, 

\item $\bm{F}_i = \bm{Z} \hat{\bm{P}}_i \bm{Z}^\top + \bm{H}_i$, 

\item $\bm{K}_i = 
    \begin{pmatrix}
    	\bm{T}_i & \bm{0} \\
        \bm{0} & \bm{T}_i
    \end{pmatrix} 
    \hat{\bm{P}}_i \bm{Z}^\top \bm{F}_i^{-1}$, 
    
\item $\hat{\bm{a}}_{i+1} =
    \begin{pmatrix}
    	\bm{T}_i & \bm{0} \\
        \bm{0} & \bm{T}_i
    \end{pmatrix}
    \hat{\bm{a}}_i + \bm{K}_i \bm{u}_i$, 
    
\item $\hat{\bm{P}}_{i+1} =
    \begin{pmatrix}
    	\bm{T}_i & \bm{0} \\
        \bm{0} & \bm{T}_i
    \end{pmatrix}
    \hat{\bm{P}}_i \left[ 
    \begin{pmatrix}
    	\bm{T}_i & \bm{0} \\
        \bm{0} & \bm{T}_i
    \end{pmatrix}
    - \bm{K}_i \bm{Z}^\top \right]^\top + 
    \begin{pmatrix}
    	\bm{Q}_i & \bm{0} \\
        \bm{0} & \bm{Q}_i
    \end{pmatrix}$, 
\end{itemize}
with the model matrices
\begin{align*}
	\bm{Z} = 
	\begin{pmatrix}
		1 & 0 & 0 & 0\\
    	0 & 0 & 1 & 0
	\end{pmatrix},\quad
    \bm{T}_i =
    \begin{pmatrix}
    	1 & (1-e^{-\beta_{s_i}\Delta_i})/\beta_{s_i} \\
        0 & e^{-\beta_{s_i}\Delta_i}
    \end{pmatrix},\quad
    \bm{Q}_i =
    \begin{pmatrix}
		\text{Var}(\xi_{ci}) & \text{Cov}(\xi_{ci}, \zeta_{ci})\\
    	\text{Cov}(\xi_{ci}, \zeta_{ci}) & \text{Var}(\zeta_{ci}) 
	\end{pmatrix}.
\end{align*}

\paragraph{Log-likelihood} Due to the assumptions of normality, the log-likelihood can be calculated as a by-product of the Kalman filter, 
\begin{equation*}
	l(\tilde{\bm{z}}_1, \dots, \tilde{\bm{z}}_n) = - n \log(2\pi) - \dfrac{1}{2} \sum_{i=1}^n 
    	\left( \log \vert \bm{F}_i \vert + \bm{u}_i^\top \bm{F}_i^{-1} \bm{u}_i \right),
\end{equation*}
where $\vert \bm{F}_i \vert$ denotes the determinant of $\bm{F}_i$.

\paragraph{Smoothing} We initialise $\bm{r}_n = \bm{0}$ and $\bm{N}_n = \bm{0}$. Then, for $i=n, n-1, \dots, 1$, the Kalman smoother equations are
\begin{itemize}
\item $\bm{L}_i = 
	\begin{pmatrix}
		\bm{T}_i & \bm{0}\\
    	\bm{0} & \bm{T}_i
	\end{pmatrix}
    - \bm{K}_i \bm{Z}$,

\item $\bm{r}_{i-1} = \bm{Z}^\top \bm{F}_i^{-1} \bm{u}_i + \bm{L}_i^\top \bm{r}_i$,

\item $\bm{N}_{i-1} = \bm{Z}^\top \bm{F}_i^{-1} \bm{Z} + \bm{L}_i^\top \bm{N}_i \bm{L}_i$,

\item $\hat{\bm{\omega}}_i = \hat{\bm{a}}_i + \hat{\bm{P}}_i \bm{r}_{i-1}$, 

\item $\hat{\bm{\Sigma}}_i = \hat{\bm{P}}_i - \hat{\bm{P}}_i \bm{N}_{i-1} \hat{\bm{P}}_i$. 
\end{itemize}

Then, $\hat{\bm{\omega}}_i$ is the smoothed state vector at time $t_i$, and $\hat{\bm{\Sigma}}_i$ the smoothed state variance matrix.

\paragraph{Missing data} The locations of the state transitions are generally not observed, and they must be treated as missing data. If the index $i$ corresponds to a missing value, the filtering equations become
\begin{itemize}
	\item $\hat{\bm{a}}_{i+1} =
		\begin{pmatrix}
			\bm{T}_{i} & \bm{0}\\
    		\bm{0} & \bm{T}_{i}
		\end{pmatrix}
 		\hat{\bm{a}}_{i}$,
	\item $\hat{\bm{P}}_{i+1} =
    	\begin{pmatrix}
			\bm{T}_{i} & \bm{0}\\
    		\bm{0} & \bm{T}_{i}
		\end{pmatrix}
    	\hat{\bm{P}}_{i} 
        \begin{pmatrix}
			\bm{T}_{i} & \bm{0}\\
   	 		\bm{0} & \bm{T}_{i}
		\end{pmatrix}^\top + 
        \begin{pmatrix}
			\bm{Q}_{i} & \bm{0}\\
    		\bm{0} & \bm{Q}_{i}
		\end{pmatrix}$,
\end{itemize}
and the smoothing equations,
\begin{itemize}
\item $\bm{r}_{i-1} = 
	\begin{pmatrix}
		\bm{T}_{i} & \bm{0}\\
    	\bm{0} & \bm{T}_{i}
	\end{pmatrix}^\top 
    \bm{r}_{i}$,
\item $\bm{N}_{i-1} = 
	\begin{pmatrix}
		\bm{T}_{i} & \bm{0}\\
    	\bm{0} & \bm{T}_{i}
	\end{pmatrix}^\top 
    \bm{N}_{i}
    \begin{pmatrix}
		\bm{T}_{i} & \bm{0}\\
    	\bm{0} & \bm{T}_{i}
	\end{pmatrix}$.
\end{itemize}
The other equations are unchanged. Note also that missing observations do not contribute to the log-likelihood. 

\newpage
\section*{Appendix S3. Posterior samples for the grey seal analysis}
Below are histograms of the posterior samples obtained for different model parameters in the grey seal case study.
\begin{figure}[htbp]
	\centering
    \includegraphics[width=0.9\textwidth]{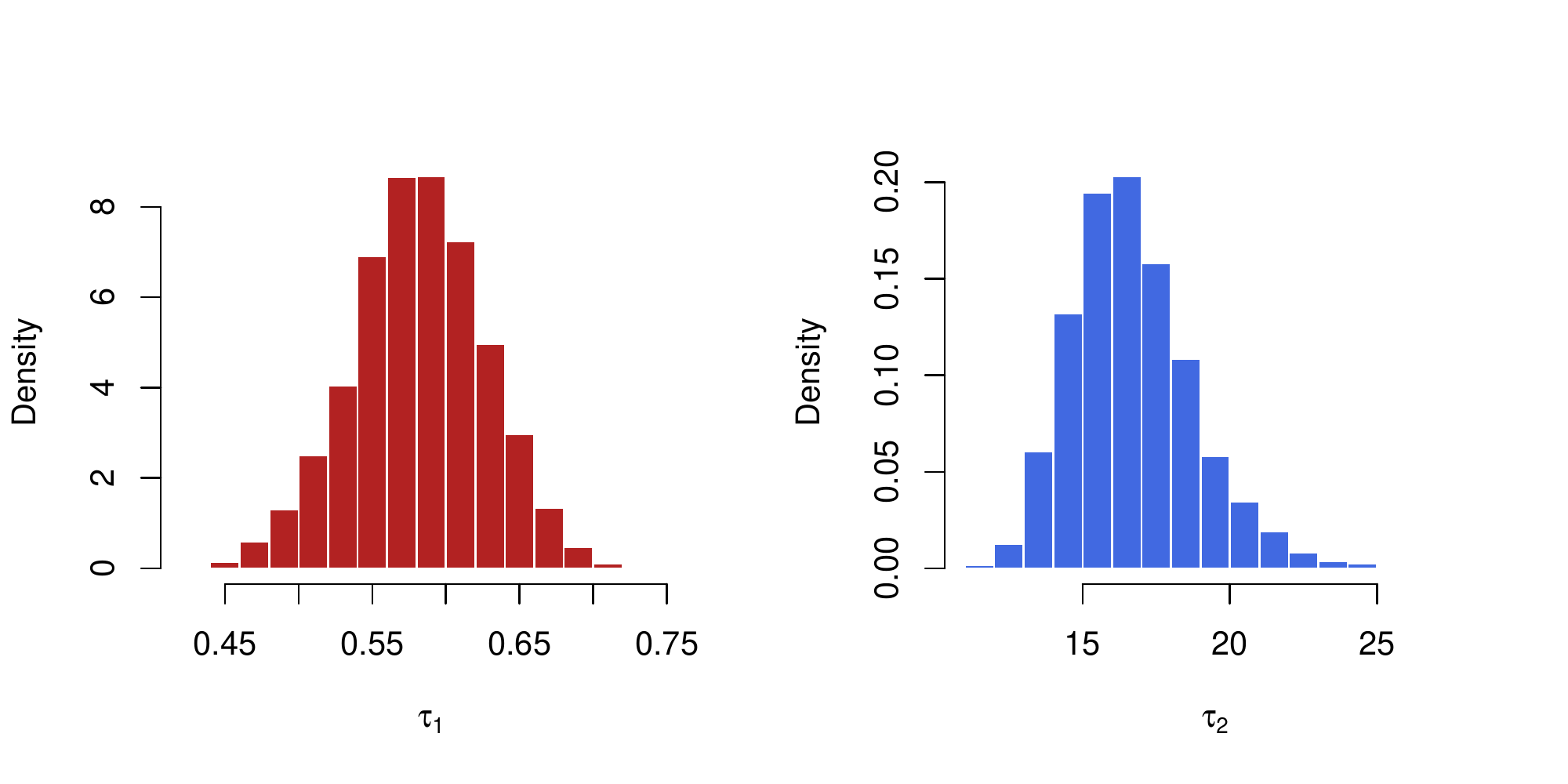}
    \caption{Histograms of posterior samples for the time scales of autocorrelation $\tau_1$ and $\tau_2$ in the grey seal case study.}
\end{figure}

\begin{figure}[htbp]
	\centering
    \includegraphics[width=0.9\textwidth]{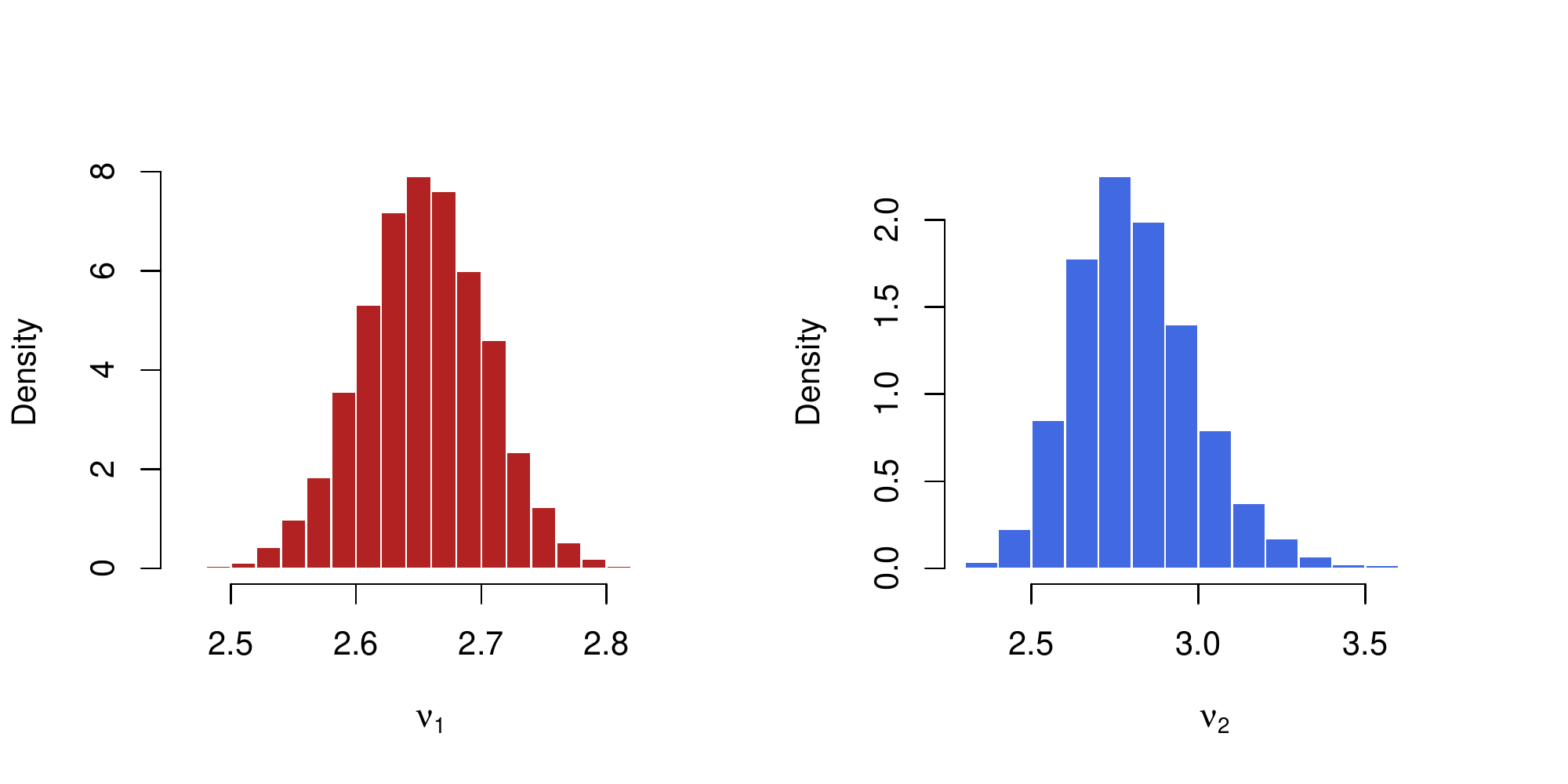}
    \caption{Histograms of posterior samples for the mean speed parameters $\nu_1$ and $\nu_2$ in the grey seal case study.}
\end{figure}

\begin{figure}[htbp]
	\centering
    \includegraphics[width=0.9\textwidth]{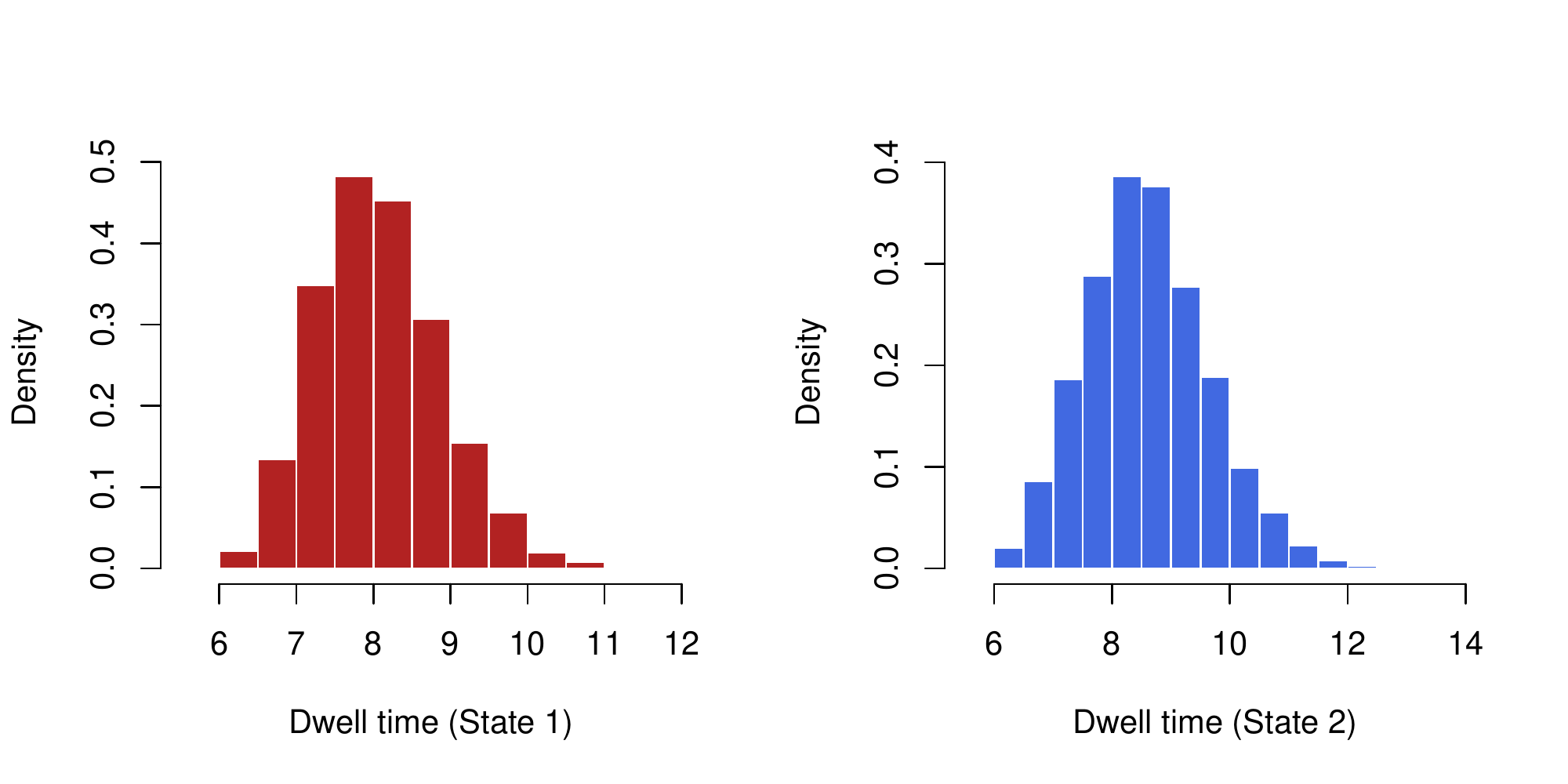}
    \caption{Histograms of posterior samples for the mean dwell times in states 1 and 2, in the grey seal case study.}
\end{figure}

\begin{figure}[htbp]
	\centering
    \includegraphics[width=0.7\textwidth]{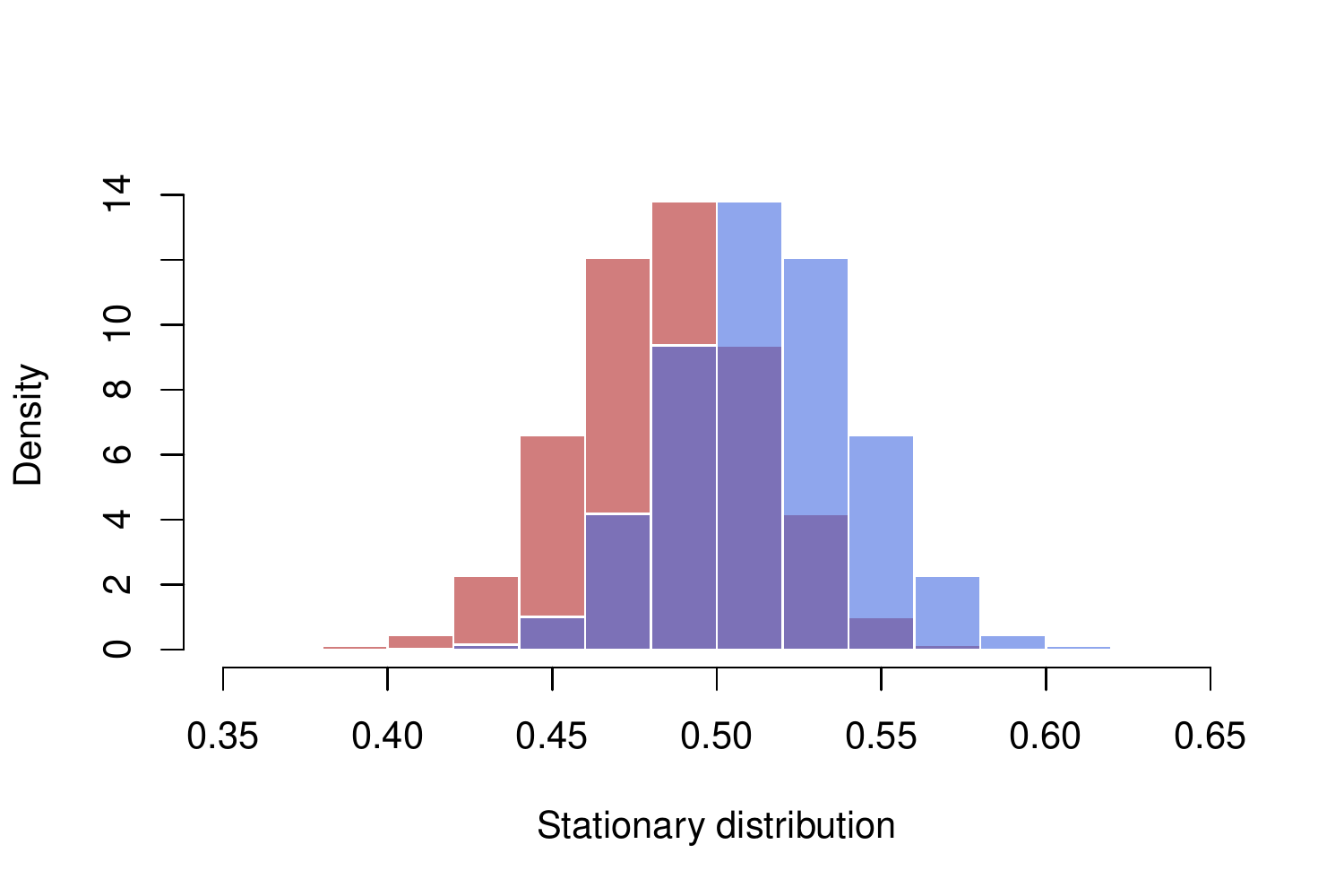}
    \caption{Histograms of posterior samples for the stationary distribution of the Markov state process, in the grey seal case study. The red histogram shows the posterior sample for $\pi_1$ (the stationary probability of being in state 1), and the blue histogram shows the posterior sample for $\pi_2=1-\pi_1$.}
\end{figure}

\end{document}